# Versatile Large-Area Custom-Feature van der Waals Epitaxy of Topological Insulators

*Tanuj Trivedi,\* Anupam Roy, Hema C. P. Movva, Emily S. Walker, Seth R. Bank, Dean P. Neikirk,\* and Sanjay K. Banerjee\**

Microelectronics Research Center, Department of Electrical and Computer Engineering, The University of Texas at Austin, Austin, TX 78758 USA

\*Corresponding Authors: tanuj@utexas.edu, neikirk@mail.utexas.edu, banerjee@ece.utexas.edu

## Abstract

As the focus of applied research in topological insulators (TI) evolves, the need to synthesize large-area TI films for practical device applications takes center stage. However, constructing scalable and adaptable processes for high-quality TI compounds remains a challenge. To this end, a versatile van der Waals epitaxy (vdWE) process for custom-feature Bismuth Telluro-Sulfide TI growth and fabrication is presented, achieved through selective-area fluorination and modification of surface free-energy on mica. The TI features grow epitaxially in large single-crystal trigonal domains, exhibiting armchair or zigzag crystalline edges highly oriented with the underlying mica lattice and only two preferred domain orientations mirrored at 180°. As-grown feature thickness dependence on lateral dimensions and denuded zones at boundaries are observed, as explained by a semi-empirical two-species surface migration model with robust estimates of growth parameters and elucidating the role of selective-area surface modification. Topological surface states contribute up to 60% of device conductance at room-temperature, indicating excellent electronic quality. High-yield microfabrication and the adaptable vdWE growth mechanism with readily alterable precursor and substrate combinations, lend the process versatility to realize crystalline TI synthesis in arbitrary shapes and arrays suitable for facile integration with processes ranging from rapid prototyping to scalable manufacturing.

## Introduction

The field of topological materials has burgeoned since the discovery of 2D and 3D topological insulators (TI),[1,2] with several prototype initial demonstrations in the offing in spintronics,[3–5] next-generation electronics,[6,7] on-chip optics and plasmonics,[8,9] and several exotic promising phenomena under intense investigation such as Majorana quantum computing,[10] axion electrodynamics and topological magnetoelectric effects.[11,12] Since the early discovery and demonstration of the staple TI compounds,[13–17] the focus of research has evolved on several fronts. Demonstrations of scalable device applications remain challenging to this day, however, with a dearth of repeatable and adaptable thin film synthesis techniques being amongst the primary reasons.[13,18] There are three well-established mechanisms to obtain high quality crystalline thin film TIs: bulk crystals and their exfoliation,[14,19–22] molecular beam epitaxy (MBE),[16,23–26] and physical vapor epitaxy,[17,27–30] also known as sub-atmospheric hot-wall van der Waals epitaxy (vdWE). The latter two are the only realistic contenders for scalable implementation. While MBE offers high quality crystalline films with a fine control over film thickness, there are limiting factors such as complexity and cost of ultra-high vacuum systems, substrate choice, difficulty of ternary/quaternary compound growth and incompatibility with high vapor pressure compounds (*e.g.*, sulfides).[31] On the other hand, vdWE offers a low-cost, facile alternative, accommodating more source, substrate, and compound thin film combinations,[32,33] but the control over film thickness and area remains challenging. An optimal balance must be achieved to explore alternatives addressing the challenges of scalability and reliability of TI synthesis for practical applications.

Selective-area growth (SAG) for compound semiconductors has received a great deal of attention owing to adaptability and ease of implementation.[34–36] SAG processes for TIs have only recently started attracting focus and the field is in its nascent stage, with proposed methods such as shadow-masked pattern and polymer imprint based local chemical modification with solvents or self-assembled molecules.[37–41] There is undoubtedly a need for fully integrable processes utilizing standard microfabrication technology to obtain large-area TI films, especially ternary and quaternary compounds, for electronic, spintronic and optoelectronic device applications. Such processes must be versatile enough to span the spectrum from academic and prototype research to scalable manufacturing. Simultaneously, unraveling the details of the growth mechanism is a necessary and significant advancement towards optimization and customization of TI SAG processes, and their extension to a larger set of compound and substrate combinations for future research and development.

As the natural next step towards technological relevance, a versatile process for large-area, crystalline TI growth in customizable features on mica is presented. The TI features grow epitaxially in large single-crystal trigonal domains of several microns in size and in any arbitrary shape of linear dimensions up to the order of 100 μm. A nonlinear thickness dependence on lateral dimensions is observed along with denuded zones at boundaries, which are explained with a semi-empirical surface migration model providing insights into the underlying growth mechanism, and the role of the selective-area surface modification. The subsequent mask layers for device fabrication can be effortlessly integrated post-growth using standard photolithography. DC transport on directly-grown TI Hall bars of different dimensions show metallic conduction down to 77 K, and the device sheet conductance remains remarkably flat with increasing TI Hall bar thickness at room temperature across several samples, indicating that the transport is dominated by the metallic topological surface states (TSS) with a low bulk contribution.

**Results and Discussion**

The custom-feature van der Waals epitaxy (CF-vdWE) growth and fabrication process is described in detail in the experimental methods section and the growth results are shown in Figure 1. The process is constructed from readily integrable steps: standard photolithography, reactive plasma etching, standard solvent cleans, and hot-wall vdWE growth of Bismuth Telluro-Sulfide ($Bi_2Te_{2-x}S_{1+x}$, $0.3 \leq x \leq 0.4$) or BTS. BTS is theoretically predicted to be one of the most promising 3DTIs to solve practical challenges of device implementation,[42] and has been shown to possess accessible TSS both from transport[17] and angle-resolved photoemission spectroscopy (ARPES) measurements.[43] The CF-vdWE process can nevertheless be easily extended to other

TI compounds in the Bi/Sb family, simply by altering the precursor material combinations in the vdWE step (see Supporting Figure S1 for examples of CF-vdWE grown $Bi_2Te_3$).

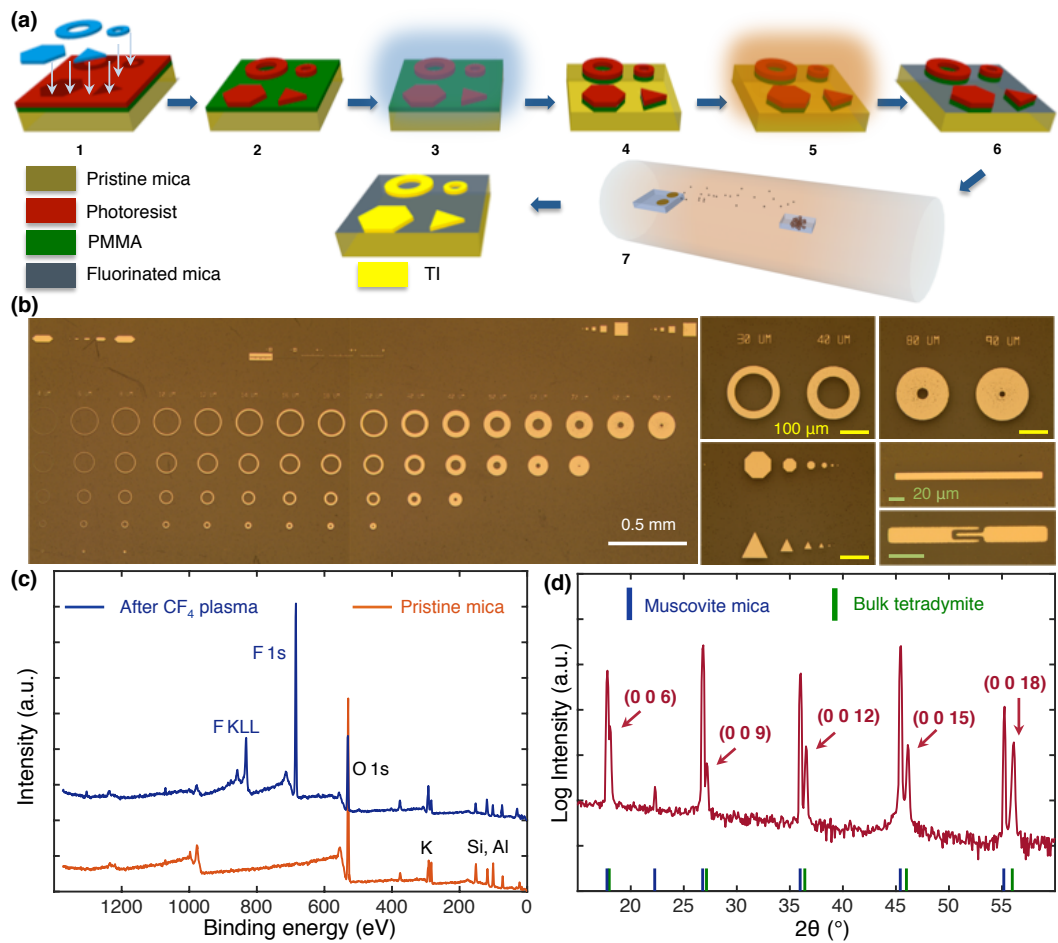

**Figure 1. Custom-feature van der Waals epitaxy process and materials characterization.** (a) Process flow schematic for custom-feature vdWE (not to scale). (b) Optical images of representative CF-vdWE grown BTS TI. (Left) Dimensionality test matrix of annuli of different widths and outer diameters, (right) CF-vdWE grown TI shapes such as annuli, hexagons, triangles, rectangular bars and a prototype microwave capacitor. (c) XPS spectra of muscovite mica substrate before (orange) and after (blue) the $CF_4$ plasma process, indicating fluorination of the surface post-process. (d) XRD pattern of CF-vdWE grown BTS. Only the (0 0 n) facet reflections of the bulk tetradymite structure are observable (green ticks), straddled next to muscovite mica peaks (blue ticks).

The fundamental process flow is schematically represented in Figure 1a. Muscovite mica is a layered inorganic compound that cleaves readily out of plane, breaking bonds at the potassium layer,[44] revealing an atomically flat and smooth single-crystal (0 0 1) plane (see Supporting Figure S2) and providing an excellent surface for TI compound growth.[30,45] While there is a large lattice mismatch (~24%) between mica (a ≈ 5.2 Å) and BTS (a ≈ 4.2 Å), layer-by-layer epitaxial growth of BTS on mica can still be obtained due to the weak substrate dependence of vdWE.[32] Supporting Figure S3 shows results of BTS growth on un-patterned pristine mica. The CF-vdWE process results in large-area contiguous BTS films highly confined within the feature boundaries, as seen in Figure 1b. The TI material grows in virtually any shape as predefined by the lithographically masked plasma process. The typical growth mask used in this experiment involves a matrix of rings or annuli of different widths (increasing from left to right) and different outer diameters (decreasing from top to bottom), as shown in Figure 1b. A variable annulus pattern matrix is chosen in order to study the dependence of the process on lateral dimensions and the pitch of an array of features, eliminating the need to pattern several different shapes with varying sizes and pitches. Remarkably, there is virtually no growth outside the feature boundaries in the $CF_4$ exposed mica regions even for growth times as long as 20 minutes, except for negligible deposition near localized physical defect sites. If the plasma process were to merely induce physical damage on the surface, the overall adhesion would be expected to improve with more growth or deposition around dislocations and defects.[46] The absence of any significant growth in areas as large as a few millimeters points to an alternative mechanism, which overcompensates for any improved adhesion. Such a mechanism must be chemical in nature, resulting in a reduction of the sticking probabilities of one or more constituent adatoms, preventing nucleation and/or compound formation. Indeed, the $CF_4$ plasma process results in a fluorination of the exposed mica surface as observed in comparative X-ray photoelectron spectroscopy (XPS) analysis shown in Figure 1c. Large F-peaks are observed in the XPS spectrum from a mica substrate following the plasma process, which are absent in the spectrum of pristine mica. The peaks do not disappear after standard cleaning or after the high-temperature furnace growth step, indicating that the surface remains fluorinated likely due to a deposition of a fluorocarbon sheath.[44,47] Pristine mica is fairly hydrophilic,[44] causing almost complete wetting of a water droplet on the surface, while the same substrate treated with a blanket $CF_4$ plasma exposure results in an increased contact angle of water (see Supporting Figure S4 for contact angle images). This is due to a reduction in the surface free energy of the fluorinated mica surface,[48] which in turn results in significant reduction in adhesion of water or the TI compound on fluorinated mica. Reduction in surface free energy due to plasma-related fluorination has been observed in several experiments.[44,47,49] Thus, highly selective growth of the TI compound is achieved, as the artificial boundary condition due to selective surface fluorination leads to an engineered surface for large-area crystalline growth well confined within the pristine mica regions.

X-ray diffraction (XRD) patterns of CF-vdWE grown BTS thin film features show very sharp peaks, appearing only at the *(0 0 n)* facet reflections of the bulk tetradymite crystal structure, as shown in Figure 1d, pointing to a highly *c*-axis oriented and layer-by-layer growth.[17] Further confirmation of crystallinity and uniformity of the TI is obtained from localized Raman spectroscopy (see Supporting Figure S5 for representative Raman spectra). Compositional analysis with XPS confirms that BTS grows within a stoichiometry range of $Bi_2Te_{2-x}S_{1+x}$, $0.3 \leq x \leq 0.4$, which is nominally dubbed the γ-phase.[17,43,50] See Supporting Section S6 for details on the compositional analysis.

AFM imaging reveals several outstanding features as shown in Figure 2. A typical AFM height profile of a section of a BTS annulus is shown in Figure 2a. The structure is composed of highly terraced single-crystal trigonal domains, extending up to several microns in lateral dimensions, which merge together to form the contiguous BTS annulus. A striking characteristic evident from AFM images is that the trigonal domains grow in one of only two orientations mirrored at 180°, suggesting an influence of the hexagonal in-plane symmetry of the underlying (0 0 1) mica surface.

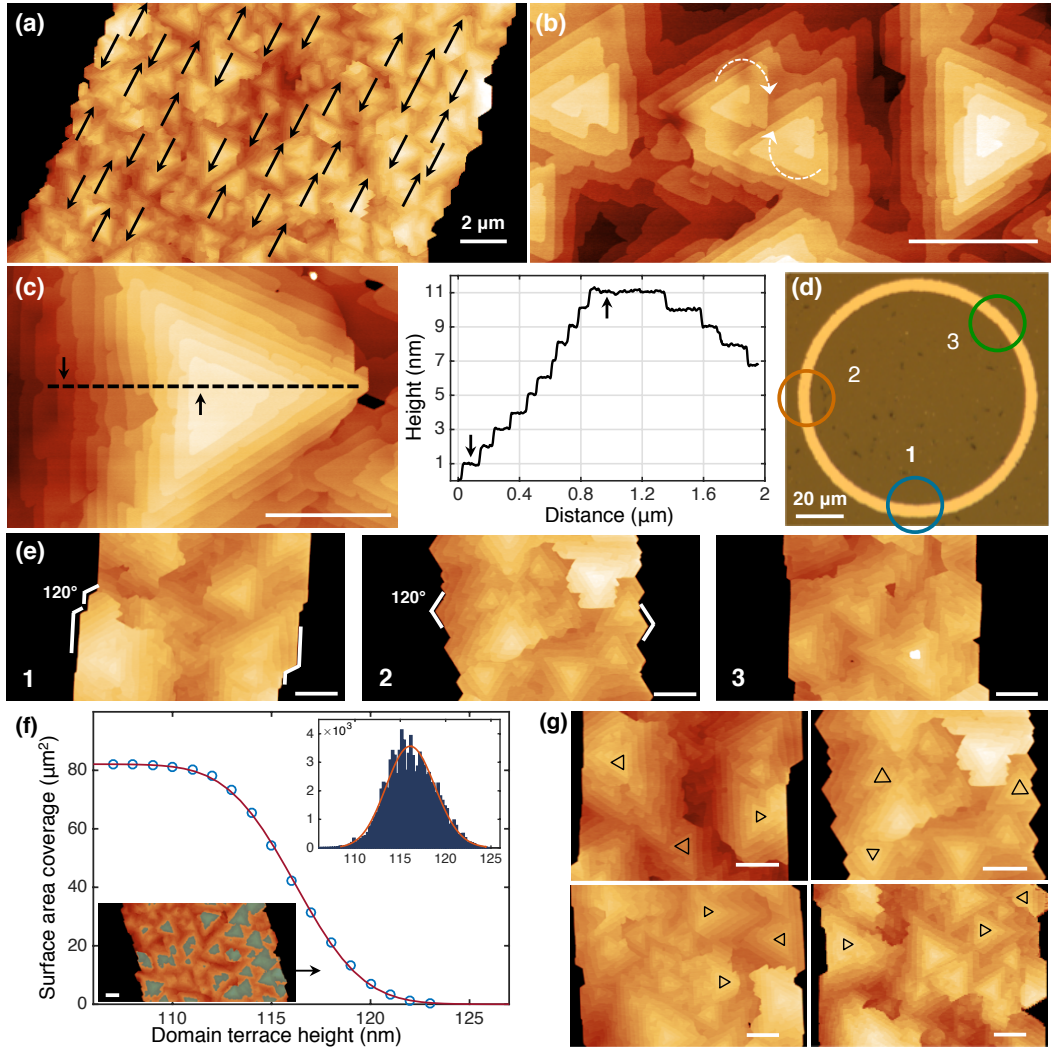

**Figure 2. Atomic force microscopy of CF-vdWE grown TI features.** (a) AFM image of a CF-vdWE grown TI annulus of 18 μm width and 200 μm outer diameter. Large, layered trigonal domains are oriented in only two directions offset at 180°. All scalebars are 1 μm unless specified. (b) A magnified AFM profile shows a cooperative spiral growth feature. (c) Magnified view of a typical trigonal domain and its height profile along the dashed line, showing subsequent layer step heights of exactly 1 nm (tetradymite crystal quintuple layer). (d) Optical image of a CF-vdWE grown TI annulus of 6 μm width. Circles indicate different locations along the perimeter. (e) Armchair-like, zigzag-like and almost straight edges of the CF-vdWE grown TI are observed depending on the location along the perimeter. (f) TI annulus surface area coverage as a function of the absolute terrace height of the constituent trigonal domains (open circles), and its lognormal CDF dependence (solid line). Top inset shows the annulus AFM thickness distribution fitted to a lognormal PDF. Bottom inset shows the area coverage at a domain height of 118 nm. (g) AFM images of CF-vdWE grown TI annuli with top-most domains indicated by black triangles, the average sizes of which are commensurate with the surface migration length of the heavier species (*i.e.,* Bi) as explained in main text.

Interesting features such as cooperative spiral growth on certain trigonal domains are also occasionally observable, as shown in Figure 2b and Supporting Figure S7. Spiral growth of trigonal terraces has been observed previously in vdWE of layered 2D materials,[33] and 3D epitaxial thin films on crystalline substrates.[51] Spiral structures typically arise as a result of screw dislocation centers propagating from the site of nucleation, providing a step source on the surface that leads to winding around the dislocation center and formation of a spiral.[51] As seen from Figure 2b, the spirals can be clockwise or counter-clockwise and can occasionally also occur as cooperative spirals. The large equilateral trigonal domains observed in the AFM images reflect the trigonal-hexagonal in-plane symmetry of the tetradymite crystal, previously observed in growths involving thin films and/or substrates with hexagonal symmetry.[31,51–53] Figure 2c illustrates a typical layered trigonal domain. The step height between each subsequent layer is approximately 1 nm, which is the thickness of one quintuple layer of the tetradymite crystal structure (see Supporting Figure S2); thus establishing that the BTS domains grow layer-by-layer in an epitaxial fashion.[17,25,26] While the edges of the TI annulus superficially appear serrated compared to the smooth lithographic boundaries in the resist, closer examination reveals highly oriented crystalline edges. AFM height profiles of the same BTS annulus at different locations along its perimeter (Figure 2d) reveal almost straight, armchair-like or zigzag-like crystalline edges exhibiting exactly 120° angles (Figure 2e and Supporting Figure S7), indicating a strong influence of relative localized orientation of the annulus perimeter with the hexagonal lattice of mica (schematically illustrated in Supporting Figure S2). Due to the artificial boundary condition, the orientation effect appears to be amplified as compared to TI growth on unpatterned pristine mica, opening up an opportunity to selectively grow thin film features in preferred orientations and with custom crystalline edges on patterned hexagonal lattices such as mica, sapphire, hexagonal BN and pyrolitic graphite. Area coverage on the surface of the CF-vdWE grown TI as a function of the absolute height of the constituent trigonal domains is shown in Figure 2f. The bottom inset shows an example of partial coverage at an absolute domain height of 118 nm, highlighted in blue. The coverage data can be accurately fitted with a lognormal complementary cumulative distribution function. Furthermore, the raw histogram data for the AFM measured thickness for the same annulus can also be fitted with a lognormal probability distribution function of the same parameters (top inset in Figure 2f). This provides further confirmation that the trigonal domains are flat and layered at steps of 1 nm. Figure 2g shows AFM height profiles of several TI annuli, indicating the topmost trigonal domains with black triangles, the significance of which will be discussed later.

Figure 3 shows a dependence of CF-vdWE grown TI thickness on the planar feature dimensions, *i.e.*, annulus width. Due to the highly layered growth, the thickness of the CF-vdWE grown TI is distributed. Figure 3a shows the evolution of the thickness distributions as a function of the annulus width from a representative growth, for a fixed outer diameter (OD) of 200 μm. As the annulus width increases, the average thickness decreases nonlinearly and shows saturating behavior, while the distributions evolve to become unimodal, exhibiting positive skewness akin to lognormal or log-logistic distributions. Figure 3b shows median thickness as a function of annulus width for four different OD sets from the same growth. The directly-grown annulus shapes conveniently provide a singular parameter (annulus width) for comparative analysis without having to find an appropriate normalization of planar dimensions of the features to their nearest-neighbor distances or pitches.[36,54] As an unusual characteristic, denuded or exclusion zones (EZ) near the feature boundaries are also observed, more evident in AFM amplitude error images. Figure 3c shows one such example, where two distinct pairs of boundaries are visible: the crystalline edges of the CF-vdWE grown TI annulus, and another smoother boundary on the outside. The external boundary is the pristine mica mesa formed during the selective-area $CF_4$ plasma process, typically 2-3 nm in height. Intriguingly, the TI domains in the central region of the patterned annulus grow and merge to form contiguous films, whereas the EZ near the feature boundary remains almost entirely denuded (schematically represented in Supporting Figure S8). In order to extract the lengths of the EZs, the two pairs of edges are extracted from the AFM image as shown in Figure 3d, and a length distribution of the difference between the two is obtained. Such distributions are shown in Figure 3e for annuli of different widths, with values centered around 150-200 nm (refer to Supporting Figure S9 for more examples).

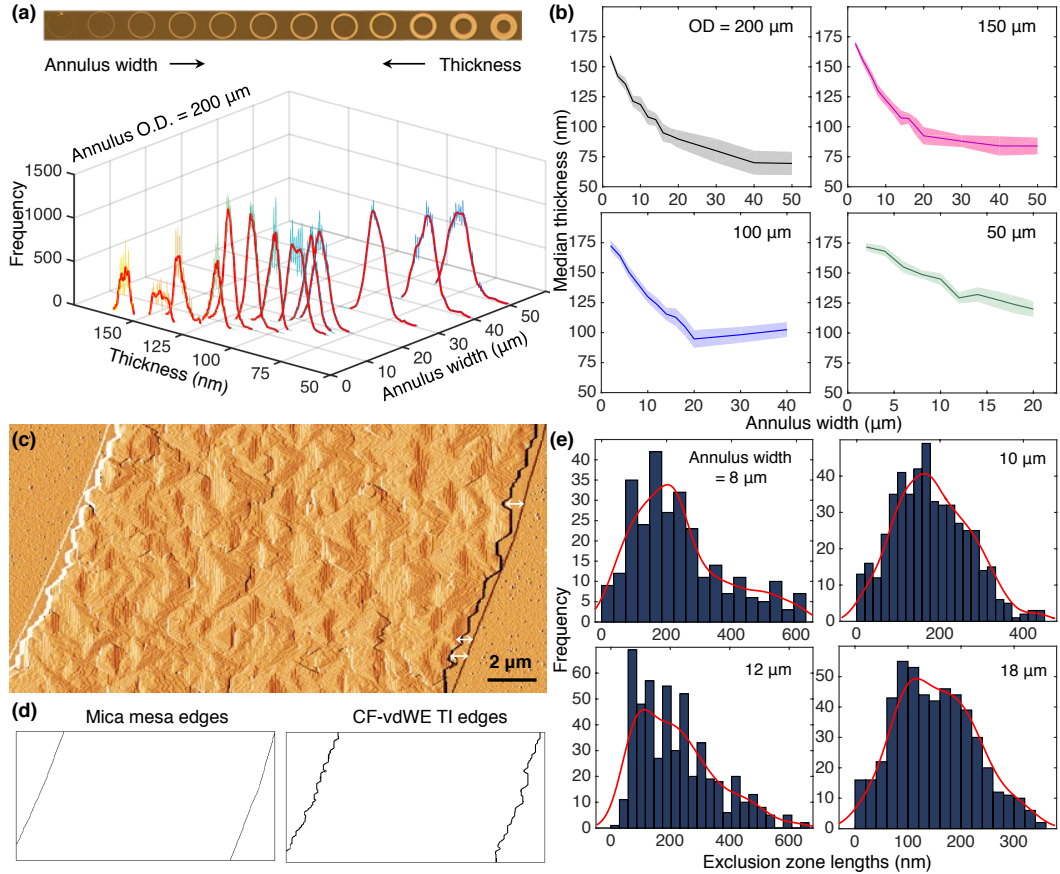

**Figure 3. Thickness variation and exclusion zones in CF-vdWE.** (a) AFM thickness distributions for CF-vdWE grown TI annuli of different widths, for an outer diameter (OD) of 200 μm. Solid lines are kernel-smoothed fits to the histograms. (b) Median thickness as a function of the annulus width for different ODs. Shaded regions represent one median absolute deviation. (c) AFM amplitude error plot of an 18 μm wide annulus, showing two distinct pairs of edges: the CF-vdWE grown TI crystalline edges and lithographically patterned pristine mica mesa edges. (d) Both pairs of edges extracted with image detection. (e) Distributions of exclusion zone lengths extracted from image detection for annuli of different widths, from the same growth. Solid lines are kernel-smoothed fits as a visual guide.

For a qualitative understanding of the underlying growth mechanism leading to the observations of an EZ and nonlinear thickness dependence, a semi-empirical two-species model is proposed. Two-species epitaxial growth modes are well studied, especially in compound systems such as GaN/As, HgTe, $Bi_2Te_3$ etc., where both species exhibit significantly different kinetic behavior on the surface during deposition and growth.[36,55,56] The custom-feature vdWE growth is largely a physical process; hence the surface migration of adatoms is expected to play a crucial role in the growth kinetics. The solid precursors $Bi_2Te_3$ and $Bi_2S_3$ incongruently sublimate to form atomic vapor fluxes, as has been observed in previous experiments.[17,27] Experimental evidence suggests that the lighter chalcogen Te and the heavier atom Bi have very different surface mobilities on mica surfaces.[57,58] Epitaxial growth studies of $Bi_2Te_3$ and related tetradymites have typically utilized Te-overpressure recipes in order to obtain high crystalline quality thin films,[31] Bi being the rate-limiter, analogous to the case of Ga in GaAs growth. However, there are important differences between the growth mechanism of MBE deposition and the custom-feature vdWE. With an initial assumption of a two-species surface migration dominated growth mechanism, we derive a simple, yet robust semi-empirical model to explain the crucial observations that render the CF-vdWE method markedly different from the case of MBE or metalorganic vapor phase epitaxy (MOVPE). The tetradymite crystal grows in a nonstoichiometric composition in reality,[50] with the S and Te atoms intermixing in the chalcogen layer of the unit cell. Moreover, the difference in surface mobility between Te–Bi and S–Bi should be of the same order, as the lighter chalcogens have comparable diffusivities in crystalline semiconductors.[59–61] Hence, a two-species model would be appropriate considering Bi as species $A$, and Te/S as species $B$. In the nominal growth condition without an artificial boundary condition as in the CF-vdWE growth, as long as the incident areal vapor flux remains constant, any two arbitrary regions of different areas should receive the same amount of flux, and hence exhibit the same thickness at the end of the growth. In order to rationalize a thickness increase for narrower annuli, an additional flux $j^{in}$ must be considered, which is dependent on the feature dimensions and can only originate from the surface diffusion of adatoms from the vast fluorinated regions surrounding the pristine mica features. The observation of an EZ near the patterned feature boundaries is also markedly different from conventional SAG experiments, where an increased thickness at abrupt boundaries is typically observed,[54] as is also observed in conventional epitaxy.[46] An imbalance in the rate of change of available adatoms near the boundary region is required for formation of an EZ, such that an impinging adatom near a feature boundary has a finite probability or rate $-J^{out}$ of escaping into the fluorinated regions without contributing to compound formation. Thus, the rationalizations that build the basis of the two-species model are: species $A$ has a significantly lower surface migration length (SML) than species $B$ on pristine and/or fluorinated mica surfaces, and that a critical imbalance exists between the additional surface diffusion flux $j^{in}$ and the rate of escape $-J^{out}$ for the formation of the EZ and increased thickness.

**Table 1.** Logical table outlining growth scenario possibilities for the two-species surface migration model

| Conditions / Observations | $N_A \ll N_B$ | | | $N_A \gg N_B$ | | | $N_A \sim N_B$ | | |
|---|---|---|---|---|---|---|---|---|---|
| | $s_A, s_B \approx 0$ | $s_A \approx 0, s_B > s_A$ | $s_B \approx 0, s_A > s_B$ | $s_A, s_B \approx 0$ | $s_A \approx 0, s_B > s_A$ | $s_B \approx 0, s_A > s_B$ | $s_A, s_B \approx 0$ | $s_A \approx 0, s_B > s_A$ | $s_B \approx 0, s_A > s_B$ |
| Thickness increase [a] | ✗ | ✗ | ✓ | ✗ | ✓ | ✗ | ✗ | ✓ | ✗ |
| Exclusion zone [b] | ✓ | ✓ | ✗ | ✓ | ✗ | ? | ? | ✓ | ? |

[a, b] Whether a thickness increase or an exclusion zone is possible, given the rate imbalance conditions explained in Supporting Information Section S10

There are a total of nine possible cases: three possible scenarios of the amount of constituent adatoms available for compound formation on the patterned mica surface, and three different scenarios of the sticking probabilities for species $A$ and $B$ on the fluorinated mica regions. These scenarios are outlined in the logical Table 1, along with the projected results from each. A satisfactory scenario that reconciles both crucial experimental observations can be arrived at by method of elimination, further described in detail in Supporting Information Section S10. The amount of effective flux contributing to growth, or number of available adatoms from the incident vapor flux for both species must be of the same order, to observe a nonlinear thickness decrease and EZ formation. The salient features of the two-species model are schematically represented in Figure 4a, where the different circles illustrate the different sticking probabilities and surface migration lengths (SML) of species *A* and *B*. An additional perimeter flux of species *B* from the fluorinated regions is represented with $j_B$, while a fractional areal escape flux of species *A* from the EZ regions is illustrated as $-J_A$. The semi-empirical model for the thickness dependence on the patterned annulus width can now be derived (see Supporting Information Section S10 for full derivation):

$$d = d_0 + \frac{\tau}{\rho_N} \cdot \frac{6 j_B \omega - 8 f J_A \lambda^2}{\omega^2 - 2\lambda\omega} \qquad (1)$$

In Equation 1, *d* is the total thickness, $\omega$ is the annulus width, $\lambda$ represents a mean exclusion zone length, $j_B$ and $f \cdot J_A$ are additional incoming perimeter flux and fractional escape areal flux for species *B* and *A*, respectively, $\tau$ is the growth time and $\rho_N$ is tetradymite number density as explained in Supporting Information. The model provides an excellent fit to the thickness dependence data as shown in Figure 4c and 4d. The extracted values of the fluxes remain virtually unchanged with growth durations or annulus OD for a given growth duration as shown in Figure 4d, indicating that the same critical rate imbalance plays a role across different growth experiments and regardless of feature dimensions. The extracted $\lambda$ from the fits also exhibit little variation as shown in Figure 4e, and are of the same order as experimentally observed EZ lengths from AFM images from Figure 3e, corroborating the validity of the two-species model. Incidentally, both the observed and extracted EZ lengths are of the same order as the size of the topmost trigonal terraces, marked as black equilateral triangles in Figure 2g. The average size of these domains is indicative of the average diffusion length or SML of the least mobile of the two adatoms, *i.e.* species *A*, on an epitaxial BTS surface. After initial nucleation at a nominally random preferred location on the pristine mica regions, a domain grows laterally and adatoms diffuse to find the lowest energy location along its edges to form the trigonal shape. For higher deposition rates, as the domain size increases, the least mobile adatoms cannot reach a domain edge quickly enough; thus formation of a new domain on the surface of the parent becomes energetically favorable.[62] While the SML of species *A* on pristine mica and the BTS surface itself should nominally be different, there seems to be a fair agreement between the values, thus providing a convenient empirical mechanism to estimate a mean SML for species *A*. The CF-vdWE growth of BTS on mica can be compared and contrasted with growth of several other technologically relevant column III/IV chalcogenide single-crystals on mica. 2D compounds like $In_2Se_3$ and GaSe exhibit layer-by-layer vdWE growth very similar to BTS on mica.[39,52] 3D materials grown on mica with vdWE exhibit contrasting growth mechanisms: such as elemental Te nanoplates that display a Volmer-Weber 3D island mode[53] and column IV chalcogenides like $Pb_{1-x}Sn_xSe$ and PbS that display 2D nanoplate growth due to lateral anisotropic mode.[63,64] In principle, the CF-vdWE growth process can be extended to grow scalable customized patterns of column III/IV chalcogenide materials on mica, for applications in on-chip photonics and optoelectronics.

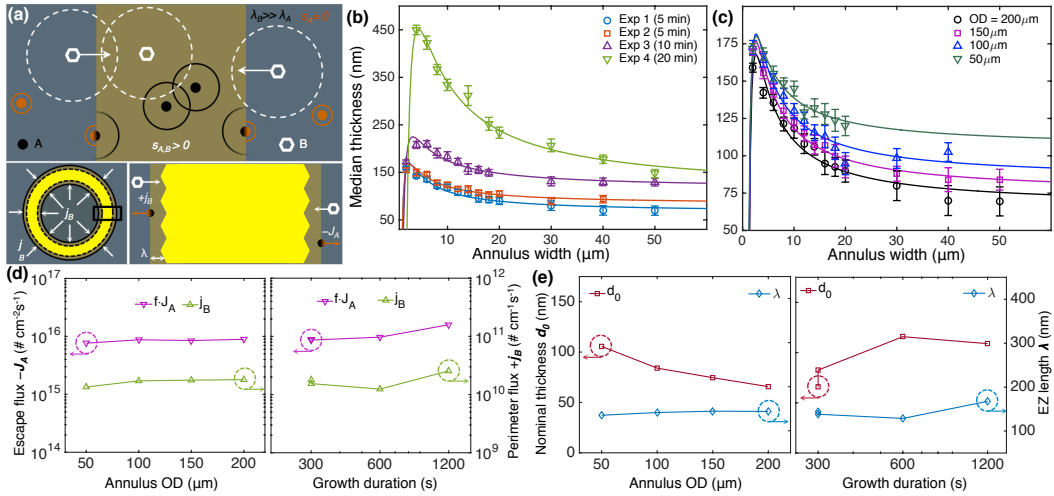

**Figure 4. Two-species surface migration growth modeling.** (a) Top schematic represents the two-species mechanism for CF-vdWE growth (not to scale). Species *A* and *B* have different sticking probabilities, and SMLs as illustrated by different circles. Bottom-left schematic shows an annulus during growth: dashed yellow annulus is the TI with a finite exclusion zone (EZ) near the feature boundary. Bottom-right schematic shows a magnified view of the black box, denoting additional perimeter flux $+j_B$ and escape area flux $-J_A$. AFM measured median thickness of CF-vdWE grown TI annuli as a function of the width from (b) different growth runs (Exp 1 – 4) and (c) different ODs from Exp 1. Solid lines are fits to the two-species model of Equation 1. (d) Extracted fractional escape flux $-f \cdot J_A$ and additional perimeter flux $+j_B$, (e) Extracted nominal thickness $d_0$ and EZ length $\lambda$, as a function of annulus OD for Exp 1, and growth durations for Exp 1 – 4.

Thus, the two-species model yields a simple and logical picture of the underlying growth kinetics due to the selective-area fluorination, without the need to numerically solve the diffusion equation, while still providing excellent empirical estimates of important growth parameters. The matrix of directly-grown annuli allows for a convenient *ex situ* mechanism for exploring growth kinetics and topographic dependence of 2D materials SAG processes in general. Different species have different surface sticking and migration behavior on fluorinated and pristine mica, which leads to selective-area growth well-confined within the feature boundaries. There is a critical flux imbalance condition that is pivotal for observing nonlinear thickness dependence and EZ formation. Further control on the thickness of the CF-vdWE grown TI can be achieved through controllably regulating the multispecies adatom flux on the fluorinated surface by changing the amount of solid precursor or the volumetric precursor flux.[45] Such a growth condition may be optimized to vary thickness across a single substrate for specialized applications, such as variable thickness grating for on-chip plasmonics and optoelectronics. Conversely, pre-patterning features of the same lateral dimensions may yield a more uniform thickness across the substrate, such that scalable TI devices can be directly grown and fabricated for applications such as spin-transfer torque memory arrays. With careful consideration of the interplay between the compound species and modified surfaces through such multi-species modeling, the CF-vdWE method can be extended to grow several different van der Waals (vdW) compounds on specifically selective-area engineered substrates.

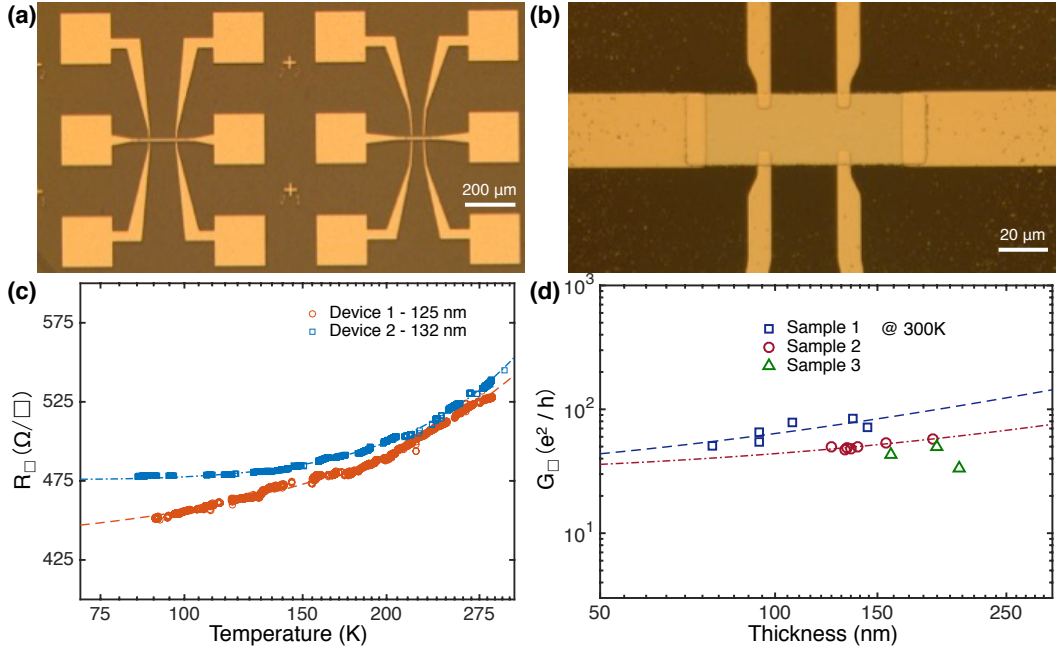

**Figure 5. DC transport on CF-vdWE grown TI devices** (a) Optical image of fully fabricated representative devices on CF-vdWE grown TI Hall bars. (b) Enlarged optical image of a typical TI Hall bar (c) Sheet resistance of two candidate CF-vdWE grown TI Hall bars as a function of sample temperature, showing a monotonic decrease in resistance, and early indications of an insulating ground state that typically manifests in TI devices at lower temperatures. (d) Sheet conductance of several different Hall bars from three different growths at room temperature, exhibiting very low bulk conductivities (150 and 61 S/cm for Samples 1 and 2, respectively). Dashed lines are fits to Equation 2.

In order to determine the quality of the TI material for electronic applications, DC transport measurements on devices of CF-vdWE grown TI Hall bars were performed, as shown in Figure 5. Due to the ease of incorporation of photolithography masks with different features into the CF-vdWE method, an array of Hall bars of variable dimensions (hence variable thicknesses) can be grown directly, and a subsequent mask can be aligned to define metallic contact leads as shown in Figure 5a and 5b. Figure 5c shows the sheet resistance of two different devices as a function of the substrate temperature, measured down to 77K in a liquid-$N_2$ probe station. Both show monotonic decrease in resistance with temperature, which is expected of the metallic nature of TSS-dominated transport in planar devices.[16,17,21,24] A reduction in the rate of decrease of resistance is observed as the temperature is decreased, which can lead to an insulating ground state at even lower temperatures, after a resistance minimum is encountered. The insulating ground state is a result of a balance between the positive conductivity contribution of the signature weak antilocaliztion (WAL) effect observed in TI devices, and the negative conductivity contribution from electron-electron interactions in the 2D Dirac fermions of the TSS manifesting at low temperatures.[17,65] Figure 5d shows the room-temperature device sheet conductance in units of $e^2/h$ as a function of the Hall bar thickness across three difference growths, exhibiting remarkably flat behavior expected from a metallic TSS-dominated transport mechanism. A two parallel channel conduction model for TI devices can be considered:[17,19,21]

$$G_{dev} = \sigma_b \cdot d + G_{ss} \tag{2}$$

In Equation 2, $G_{dev}$, $\sigma_b$ and $G_{ss}$ are the total device sheet conductance, bulk conductivity and surface state conductance, respectively and $d$ is the TI thickness. This model considers an effective TSS conduction channel $G_{ss}$, while other parasitic contributions such as bulk conduction due to native defects and chalcogen deficiency doping,[17,42] and elastic scattering between the bulk and TSS channels, can be lumped into an effective contribution $\sigma_b$. The linear fit of Equation 2 is applied to the experimentally measured $G_{dev}$ for Sample 1 and 2, to extract the $\sigma_b$ values of 150 S/cm and 61 S/cm, respectively, signifying very low bulk conduction that is comparable to bulk-insulating exfoliated BSTS devices,[21] likely due to lower bulk defects and chalcogen deficiencies. The fits also yield the y-axis intercept for Sample 1 and 2, i.e. $G_{ss}$, as 23.7 and 28 in units of $e^2/h$, respectively, indicating similar 2D TSS metallic conductivity and uniformity across devices from separate growths. For the devices of Sample 1 and 2, the contribution of the 2D Dirac TSS to total conduction at room-temperature is scattered around 50% (Supporting Figure S11) with the largest one being at 60%, which is amongst one of the highest reported room-temperature conduction ratios in synthesized TI thin films, rivaling that of bulk crystal devices of BSTS.[20,22] Excellent transport and optical properties of TSS for devices of comparable thicknesses have been previously reported for epitaxial thin films and bulk crystal exfoliated flakes for high quality crystalline TIs.[5,8,9,16,24] At lower operating device temperatures, imperative for several TI applications involving proximity-effect heterostructures with superconductors and ferromagnets, the TSS contribution is expected to increase as the bulk carriers are frozen out, further improving the device characteristics. Moreover, due to the highly crystalline, chemically inert and insulating mica bottom interface, substrate related scattering limiting TSS mobility is expected to be negligible.[66] The DC transport measurements establish a TSS-dominated conduction mechanism in the directly-grown TI devices, with a promisingly low bulk contribution and an intrinsic chemical potential at room-temperature. The high quality CF-vdWE grown TI shows great potential for implementing practical devices on large-area crystalline arrays for applications such as in spin-based memory and logic,[3,5,7] and on-chip optics and plasmonic devices.[8,9]

## Conclusions

In conclusion, a scalable and high-yield custom-feature vdWE method using selective-area surface modification through microlithographically masked fluorination is presented for realizing large-area crystalline growth of TI compounds on mica. Large terraced single-crystal trigonal domains are observed, which merge to form contiguous thin films. The features exhibit a highly oriented growth with the underlying hexagonal mica lattice, uncovering the prospect of growing TI and 2D materials in preferential orientations on specifically engineered vdW substrates. The thickness of the CF-vdWE grown TI has a nonlinear dependence on the planar feature dimensions, which can be described well by a semi-empirical model considering two-species surface migration on the mica surface. Transport measurements on CF-vdWE grown TI Hall bars reveal TSS-dominant conduction with low bulk conductivity, indicating excellent electronic quality for on-chip applications involving probing and manipulation of the TSS. The CF-vdWE method can be readily extended to wafer-scale large area crystalline TI growths. The vdWE method additionally provides a facile way to exchange source precursors with minimal alteration to introduce dopants or different compound combinations, to grow a plethora of layered 3DTI compounds from the tetradymite family, i.e., $(Bi_ySb_{1-y})_2(Te_{1-x}\{Se/S\}_x)_3$. In principle, this method also presents a promising candidate for exploring custom-feature large-area growth of other technologically relevant 2D vdW materials such as transition metal and column III/IV chalcogenides for next-generation electronics and photonics applications. The CF-vdWE process achieves a versatile growth method harnessing planar microfabrication processes to obtain large-area crystalline TI structures for electronic, spintronic and on-chip optical device applications, while simultaneously being highly adaptable to prototype research as well as optimized scalable implementation.

## Experimental Section

*Lithographic modification of mica substrates*: The fabrication and growth process for the custom-feature TI growth on pre-patterned mica substrates is schematically represented in Figure 1a. Muscovite mica disks of 10-25 mm diameter (Ted Pella Inc.) were cleaved along the (0 0 1) plane immediately prior to the process using a clean scalpel. A layer of PMMA A4 (MicroChem) was spin-coated at 4k rpm on the freshly cleaved substrates and baked at 180 °C, followed by a layer of AZ 5209E photoresist (PR) spin-coated at 4k rpm, baked at 90 °C. A mask aligner with an i-line UV source at 7.5 mW cm$^{-2}$ intensity was used to expose a custom-designed pattern from a photomask onto the mica substrate with the dual-resist layers in vacuum contact mode (Step-1). The PR layer was then developed using a standard 2.3% tetramethylammonium hydroxide (TMAH) developer (Dow MF-26A) (Step-2). As the cleaved muscovite mica surface contains Al and Si oxides, it reacts with TMAH if exposed directly and is slowly etched, leading to low-yield in a single layer resist process. The PMMA layer, which is inert to TMAH, protects the mica surface during development and prevents unexposed PR from peeling off. The substrates were then loaded into an RIE plasma chamber (Plasmatherm 790) for a dual-step plasma process: (1) A 100 W oxygen plasma to transfer the patterns from the PR to the PMMA film underneath (Step-3, 4), (2) Without breaking vacuum, a 100 W $CF_4$ plasma to fluorinate the exposed mica surface (Step-5, 6). Test mica substrates without any lithographic patterns were also loaded into the RIE chamber, to be used later for contact-angle measurements. The substrates were then cleaned in hot NMP (Remover PG, MicroChem) overnight to remove resist and other organic contaminants.

*vdWE growth and materials characterization*: The cleaned fluorinated mica substrates were loaded into the vdWE growth furnace. Detailed description of the growth system and method can be found elsewhere.[17] The precursor materials in the central zone were ramped up to 510 °C, such that the sublimated vapor flux is carried over to a cold zone of the furnace by an inert carrier gas ($N_2$), where the pre-patterned clean mica substrates were horizontally arranged. The substrate temperature was typically in the range of 390 – 410 °C, the chamber pressure was maintained at 20 – 50 Torr and the $N_2$ gas flow rate was typically 100 – 150 sccm. The central zone temperature was held constant typically for 5 – 20 minutes, before cooling down naturally to room temperature (Step-7). The composition post-growth was confirmed by XPS analysis (SCALAB Mark II Omicron) on the mica substrates. Sample-wide crystallinity of the CF-vdWE grown features was determined with XRD (Philips X'Pert) and locally with scanning Raman spectroscopy (Renishaw inVia). An in-house goniometer with a digital camera was used for measuring the contact angle of water on test mica substrates before and after the $CF_4$ plasma process to establish the surface free energy difference. Tapping mode AFM (Veeco Nanoscope V) was used to extensively image the grown features locally, and to extract thickness distributions, domain sizes and orientations, and exclusion zone boundaries. Statistics, image analysis and fitting were performed with MATLAB. Open source SPM software Gwyddion was utilized for processing acquired AFM data.[67]

***Device fabrication and transport measurements:*** A lithographic mask that has rectangular bars of different dimensions was used to pre-pattern mica substrates and directly grow TI bars from scratch. After growth, a second mask layer comprising of the contact leads and pads was aligned on top of TI features using a similar double-resist layer photolithography process as before. Immediately prior to metallization, the developed contact regions were exposed to a brief Ar RIE plasma process to remove surface oxides and to improve contact adhesion. Subsequently, a metal stack of Ti/Pd or Ti/Au was deposited using e-beam evaporation, with typical metal thicknesses in the range of 3 – 5 nm Ti and 120 – 150 nm Pd or Au. Samples were then placed in hot NMP overnight for liftoff. The DC transport measurements were performed with either Cascade Microtech Summit probe station in air at room-temperature or Lakeshore FWPX cryogenic probe-station down to liquid nitrogen temperatures in vacuum, using SRS-830 lock-in amplifier or the Agilent B1500 semiconductor parameter analyzer.

## Acknowledgments

This research was supported in part by the Semiconductor Research Corporation's NRI SWAN program and the NSF National Nanotechnology Coordinated Infrastructure (NNCI).

The authors declare no competing financial interest.

**Supporting Information**

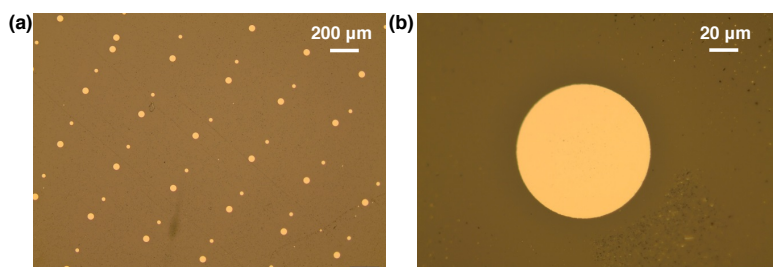

**Figure S1.** Optical images of custom-feature van der Waals epitaxially (CF-vdWE) grown $Bi_2Te_3$ circles on selectively fluorinated mica substrates.

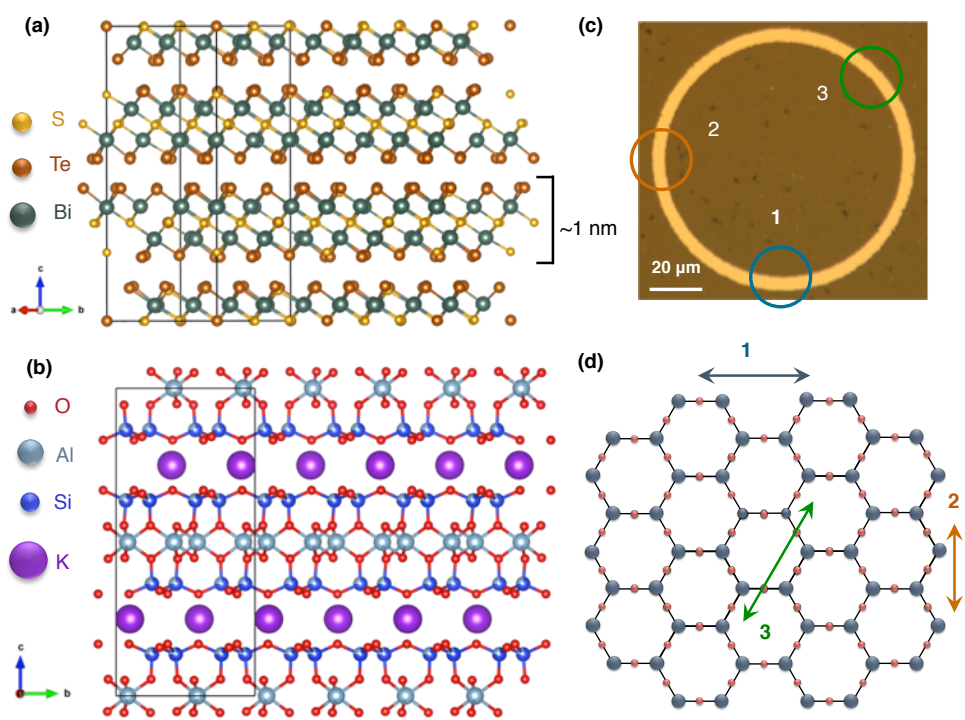

**Figure S2.** (a) Schematic view of the layered Bismuth-Telluro Sulfide (BTS) crystal structure, with the unit cell outlined in black. A quintuple layer of thickness 1 nm is also illustrated. (b) Schematic view of the layered muscovite mica crystal structure, with its unit cell outlined in black. (c) Optical image of a CF-vdWE grown BTS annulus with three different locations highlighted on its perimeter. (d) Schematic view of the muscovite mica (0 0 1) in-plane hexagonal lattice, which forms the growth substrate. The orientations of the three locations along the annulus perimeter from (c) are illustrated on the underlying hexagonal lattice, which enforces armchair and zigzag crystalline edges on the BTS annulus. Crystal structure schematics were generated with open-source software VESTA 3[S1] using data from American Mineralogist open database.[S2]

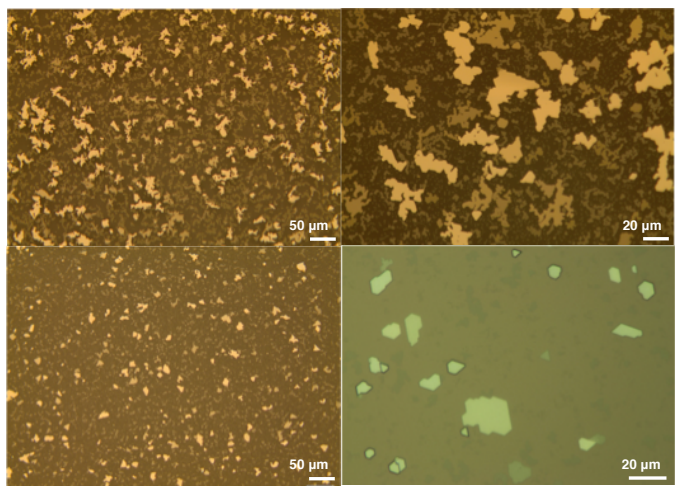

**Figure S3.** Optical images of vdWE grown BTS on unpatterned mica substrates.

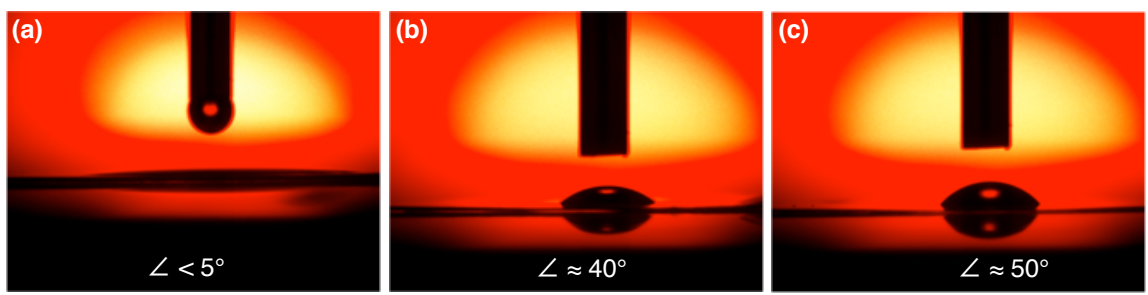

**Figure S4.** Optical images showing contact angle measurements of water droplets on muscovite mica. (a) Pristine mica substrate, (b) after 50 W $CF_4$ RIE process (c) after 100 W $CF_4$ RIE process. Increasing contact angle indicates reduced surface free-energy of the fluorinated mica surface.[S3,S4]

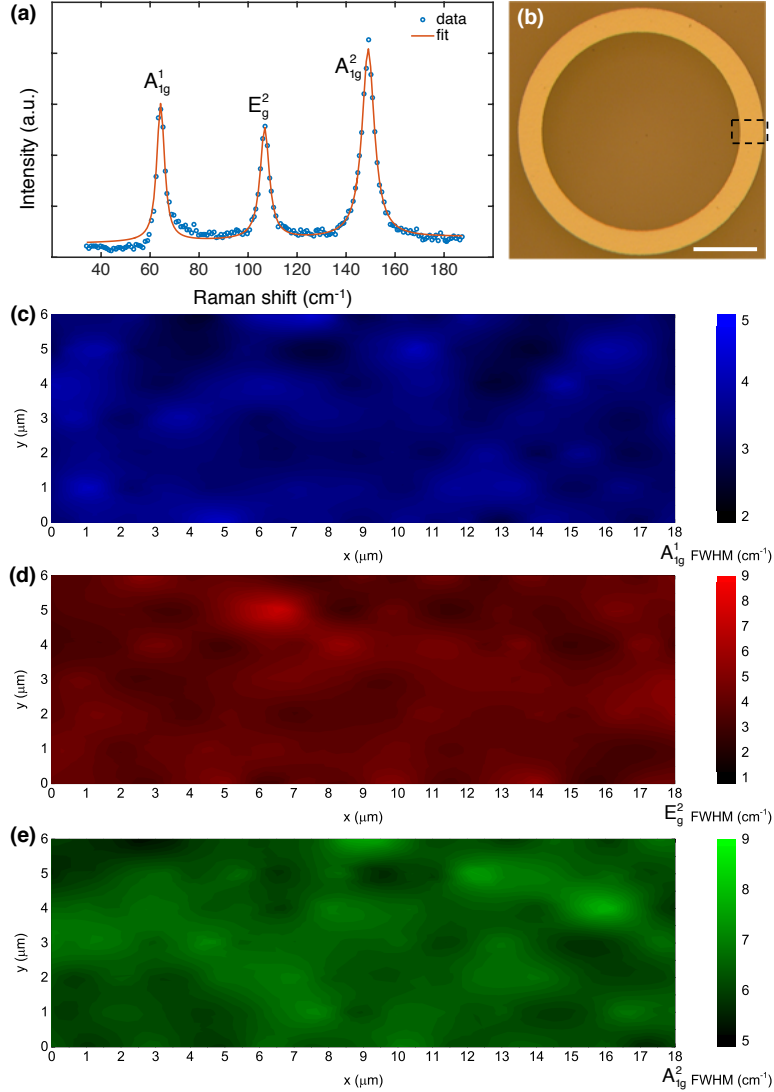

**Figure S5.** (a) Representative Raman spectrum of CF-vdWE grown BTS annulus showing three sharp peaks and their fits, blue-shifted from that of $Bi_2Te_3$.[S5–S7] (b) Optical image of a BTS annulus, showing the region where 2D mapped Raman spectra were measured with a dashed black box (scale bar is 50 μm). Smoothed XY maps of fitted full-width half-max (FWHM) for (c) $A^1_{1g}$, (d) $E^2_g$, and (e) $A^2_{1g}$ peaks, showing good uniformity.

## Section S6 Compositional analysis of CF-vdWE grown BTS

The composition of the CF-vdWE grown BTS films across different growth runs remains within the narrow range of stoichiometry reported in the main text. The BTS compound, also referred to as tetradymite, has been previously largely explored only from bulk crystal growths. The data in literature on synthetic crystals of tetradymite is limited, main works being those of Glatz (1967),[S9] Soonpaa (1962)[S10] and more recently, Ji *et al.* (2012).[S11] Glatz thoroughly examined the $Bi_2Te_3$ – $Bi_2S_3$ system with bulk crystal growths, and Pauling subsequently analyzed Glatz and Soonpaa's work by theoretical arguments on the structure of the tetradymite.[S12] Glatz observed two compound phases of tetradymite, β-BTS and γ-BTS, in the range of 25-30% and 34-50% mole fraction $Bi_2S_3$ in $Bi_2Te_3$, respectively. Pauling calculated the stoichiometry of β-BTS as $Bi_{14}Te_{15}S_6$ and γ-BTS as $Bi_{14}Te_{13}S_8$. Pauling also observed that the ideal stoichiometry of $Bi_2Te_2S$ would require a mole-fraction of 33.3% $Bi_2S_3$, which was not observed in Glatz or Soonpaa's experiments. Ji et al.'s recent experiment investigating BTS with ARPES and our previous magnetotransport study of vdWE grown BTS nanosheets on $SiO_2$, also obtained the Sulfur-rich γ-BTS compound.[S5] The literature seems to suggest the stability of the γ-BTS compound over that of the other possible phases, and over the perfect stoichiometry of 2:2:1. According to Glatz's phase diagram of the $Bi_2Te_3 - Bi_2S_3$ system, he reported limited solubility of $Bi_2S_3$ in $Bi_2Te_3$ referred to as the α-phase. This phase extends up to only 4% mole fraction of $Bi_2S_3$ and was not observed below the solidus, and similarly was not observed in BTS nanosheet growth on $SiO_2$ or mica. This phase diagram is helpful in explaining the observations of CF-vdWE grown BTS samples, where the growth temperature range is 390 – 410 °C. These growth conditions promote the tetradymite compound growth in the S-rich γ-BTS phase. Due to a higher vapor pressure, there is likely to be a large amount of Sulfur flux ever present on the substrate during growth. The material analysis in this work, including global XRD peaks matching closely with the γ-BTS phase as reported by Ji et al.[S11] and the previously reported BTS nanosheets,[S5] confirms that BTS grows in the narrow range of stoichiometry $Bi_2Te_{2-x}S_{1+x}$, $0.3 \leq x \leq 0.4$ of the γ –phase (within instrument sensitivities).

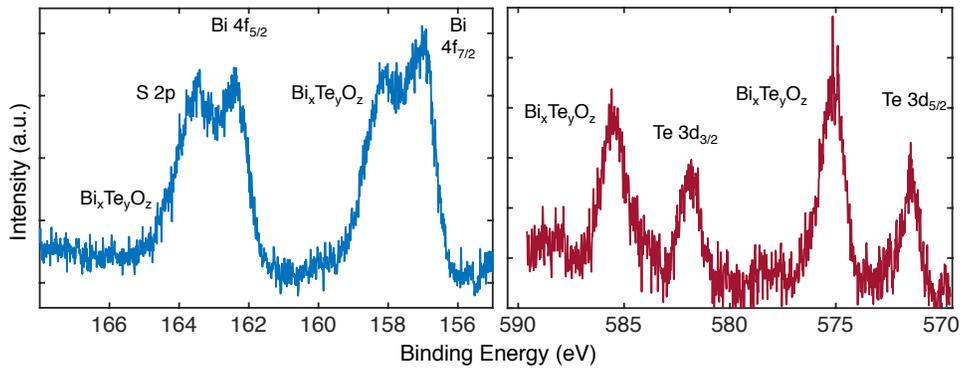

**Figure S6.** X-ray photoemission spectra (XPS) of a representative CF-vdWE grown BTS sample showing the (left) Bi, S peaks and (right) Te peaks. Surface oxidation related peak splitting is visible, as reported previously.[S5,S8]

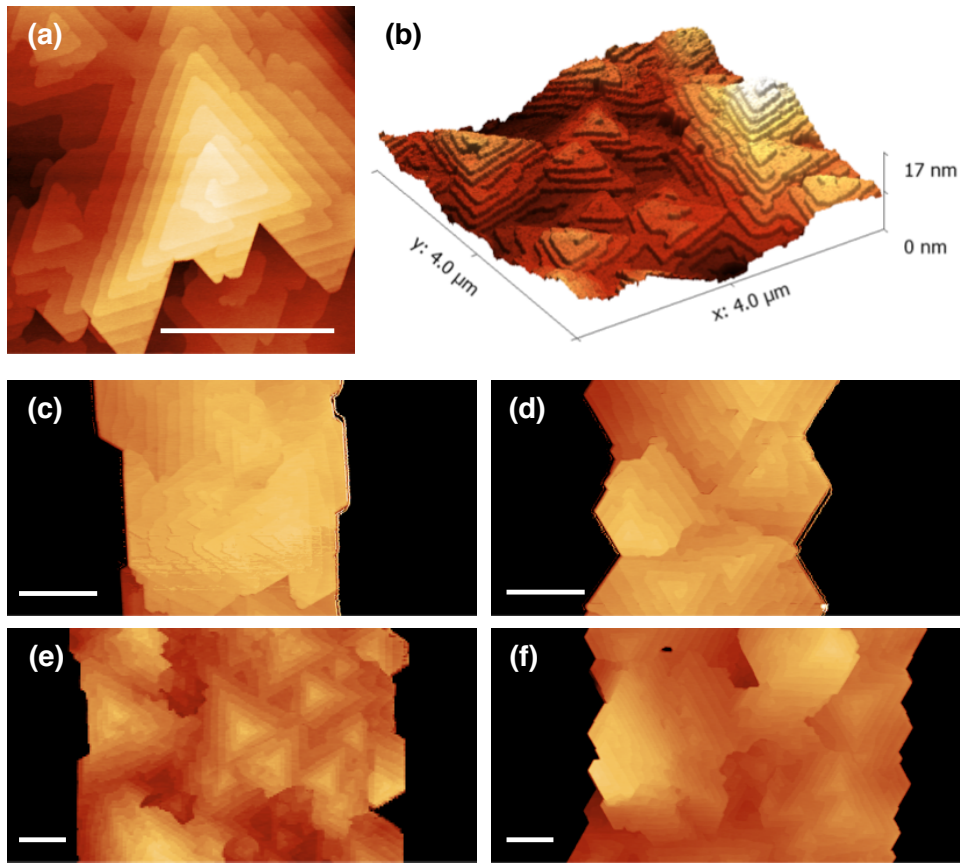

**Figure S7.** (a) Example of a cooperative spiral growth on a CF-vdWE grown BTS trigonal island. (b) 3D AFM height profile of a section of CF-vdWE grown TI annulus showing highly layered growth. Examples of oriented crystalline edges in TI annuli: (c) armchair and (d) zigzag edges in a 4 μm annulus, (e) armchair and (f) zigzag edges in an 8 μm annulus. All scalebars are 1 μm.

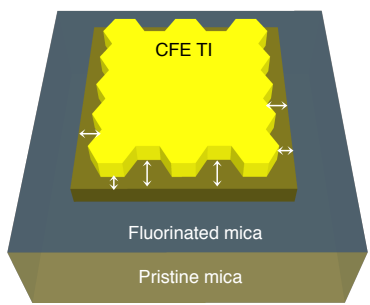

**Figure S8.** Schematic of exclusion zone (EZ) formation on selectively fluorinated mica substrate during CF-vdWE TI growth (not to scale). The EZ lengths are illustrated as white arrows.

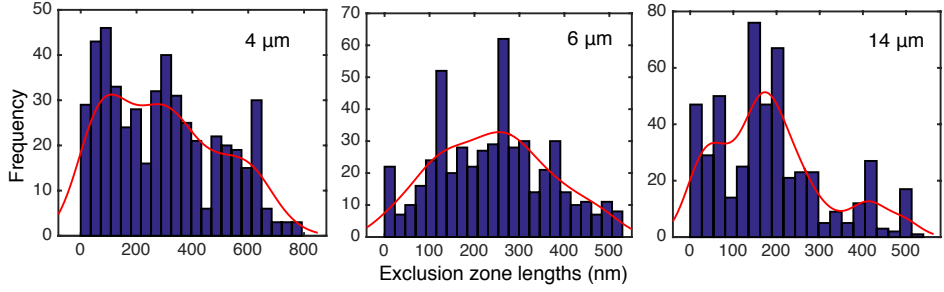

**Figure S9.** More examples of EZ length histograms for CF-vdWE grown TI annuli of different outer diameters (OD), extracted by edge-detection from AFM amplitude error images, as illustrated in main text Figure 3c and 3d.

## Section S10 Derivation of the two-species surface migration growth model

The motivation and introduction for the two-species surface migration model are presented in the main text, with the schematic and the model fits shown in Figure 4. The derivation for the fit describing the thickness dependence on annulus width is presented in this section, considering a satisfactory growth scenario that can be arrived at by a method of elimination for those listed in the logical Table 1 in detail in main text.

As outlined in Table 1, there are three possible scenarios of the amount of constituent adatoms available for compound formation on the patterned mica surface, and three different scenarios of the sticking probabilities for species $A$ and $B$ on fluorinated mica. As an example, consider the scenario where the effective available flux or number of adatoms on the patterned mica mesa has a large imbalance such that $N_A \ll N_B$, which is the typical situation in the case of MBE growth of $Bi_2Te_3$. There are three further possibilities: (1) either both species have zero sticking ($s_A$, $s_B \approx 0$), or (2) $A$ has zero sticking probability on fluorinated mica, while for $B$ it is finite ($s_A \approx 0$, $s_B > s_A$), or (3) vice versa ($s_B \approx 0$, $s_A > s_B$). For the first condition, an EZ can form near the feature boundaries, if the escape rate of $A$ is faster compared to the rate of additional influx of $B$ over an average distance of the order of $\lambda_A$ (surface migration length or SML of species $A$) from the boundary, leading to a critical imbalance of $N_A$ in $N_{AB}$ to significantly reduce compound formation near the boundary. However, a nonlinear thickness increase is not possible, as there cannot be an extra influx of atoms diffusing in from the fluorinated surfaces. For the second (third) condition, the additional flux $\Delta j^{in}$ consists of extra $B$ ($A$) adatoms diffusing inwards from the fluorinated regions, and $-\Delta j^{out}$ consists of $A$ ($B$) adatoms near the feature boundary escaping into the fluorinated regions to almost instantly desorb. In the second condition ($s_A \approx 0$, $s_B > s_A$), an EZ is possible at the boundary, similar to the first condition. An EZ cannot form for the third condition ($s_B \approx 0$, $s_A > s_B$) however, where species $A$ diffuses inwards from the fluorinated regions. The additional incoming $A$ adatoms can form a compound near the boundary irrespective of a rate imbalance, as the overall concentration of species $B$ is significantly higher than $A$. On the other hand, a thickness increase is not possible in the second condition, as a reduction in effective $N_A$ leads to a direct reduction in total moles of compound molecules $N_{AB}$, notwithstanding additional $B$ adatoms. Nevertheless, both the thickness increase and EZ formation cannot occur such that neither of the scenarios represented by the three conditions can completely describe the underlying growth mechanism behind the experimental observations. Similar arguments can be extended for the other cases enumerated in the table, until a satisfactory condition is found through the method of elimination. Such a condition requires the concentration of species A and B to be of the same order, and $s_A \approx 0$, $s_B > s_A$ on fluorinated mica.

Consider the area of incident areal vapor flux, directly contributing to the growth (explicitly omitting the exclusion zone (EZ) boundaries):

$$\begin{aligned} A^{in} &= \pi\left(R^2 - r^2\right) - \pi\left(R^2 - (R-\lambda)^2\right) - \pi\left((r+\lambda)^2 - r^2\right) \\ &= \pi\left(R^2 - r^2\right) - 2\pi\lambda(R+r) \\ &= \pi(R+r)(\omega - 2\lambda) \end{aligned} \quad (S1)$$

Where $R$ and $r$ are the outer and inner diameter of the annulus, respectively, and $\omega = R - r$, which is the annulus width. $\lambda$ represents an average value of the EZ size. The evolution of the two species $A$ and $B$ can be expressed as follows, where the superscripts $i$ and $f$ denote initial and final quantities, respectively:

$$\begin{aligned} &2\,A + 3\,B \rightarrow 1\,A_2B_3 \\ &N_A^f = N_A^i - \Delta N_A,\; N_B^f = N_B^i + \Delta N_B \\ &N_{AB}^f = N_{AB}^i + \Delta N_{AB},\; \Delta N_{AB} > 0 \text{ iff } \Delta N_A < \tfrac{3}{2}\Delta N_B \end{aligned} \quad (S2)$$

Equation S2 describes the condition in which a thickness increase may be possible as long as the inequality is satisfied. This condition is also conducive to formation of an EZ, as long as there is a steady outflux of $A$ in an average radius of the order $\lambda_A$ (surface migration length or SML of species $A$) near the feature boundary, leading to a local reduction in $\Delta N_{AB}$. The additional change in the moles or number of atoms ($\Delta N_{AB}$) of the compound can only be positive if the final inequality in Equation S2 can be satisfied. The individual species changes can be expressed in terms of a rate of change of the species normalized to spatial dimensions, i.e., a flux-like quantity. As illustrated in the schematic of Figure 4a, the additional surface influx of species $B$, $+j_B$, can be considered as a "perimeter flux" that enters from the both the inner and outer perimeters of the annulus under consideration, and hence depends on the geometry of the annulus. The escape flux of species $A$ localized near the EZ can be represented as a fraction $f \cdot \eta_A$ of the total incident areal vapor flux of species $A$, or $-f \cdot \eta_A \cdot J_A$. $\eta_A$ is the ratio of the area of the EZ[1] to the total patterned annulus area. The area ratio $\eta_A$ provides the number of species $A$ adatoms that fall just in the EZ near the boundary, and a further fraction $f$ of that number may escape to the fluorinated regions. Then the change in the species concetration in Equation S2 can be written in terms of the additional fluxes as:

$$\Delta N_{AB} = 2\Delta N_A + 3\Delta N_B = (3j_B\tau)\cdot 2\pi(R+r) - (2f\eta_A J_A \tau)\cdot 2\pi\lambda(R+r) \quad (S3)$$

---

[1] The area of the EZ is the vanishingly narrow annuli formed between the CF-vdWE TI crystalline edges and the patterned annulus boundary, which is the second term in the second line of Equation S1

In Equation S3, $\tau$ is the growth duration, $j_B$ and $J_A$ are the species **B** additional perimeter flux and species **A** escape area flux, respectively, and $\eta_A$ represents the fraction of the species **A** in the EZ area that escape to the fluorinated regions. The total number of available adatoms for growth can be expressed as:

$$N_{AB}^f = N_{AB}^i + \Delta N_{AB}$$
$$\frac{N_{AB}^f}{\rho_N} = \frac{N_{AB}^i}{\rho_N} + \frac{\Delta N_{AB}}{\rho_N} \Leftrightarrow V_{AB}^f = V_{AB}^i + \frac{\Delta N_{AB}}{\rho_N} \quad \text{(S4)}$$
$$d_{tot} \cdot A^{in} = d_0 \cdot A^{in} + \frac{\Delta N_{AB}}{\rho_N}$$

In Equation S4, $\rho_N$ is the number density of the BTS compound and $V^{f,i}$ are the volume of the feature in the final and initial conditions, respectively. The total thickness of the feature is $d_{tot}$, the nominal thickness considering only an incident areal vapor flux is $d_0$ and $A^{in}$ is the active area directly contributing to the growth, from Equation S1. Equation S4 can be rewritten from Equations S1 and S3:

$$d_{tot} = d_0 + \frac{1}{A^{in}} \cdot \frac{\Delta N_{AB}}{\rho_N} = d_0 + \frac{2\pi(R+r)}{\pi(R+r)(\omega - 2\lambda)} \cdot \left[ 3\frac{j_B \tau}{\rho_N} - 2\lambda \eta_A \frac{f J_A \tau}{\rho_N} \right]$$
$$= d_0 + \frac{\tau}{\rho_N} \cdot \frac{6 j_B \omega - 8 f J_A \lambda^2}{\omega^2 - 2\lambda \omega} \quad \text{(S5)}$$

Parameterized fits of Equation S5 to AFM measured median thickness data can be performed as a function of the annulus width; the results of which are shown in main text Figure 4c and 4d. The fit yields all four unknowns, an initial thickness $d_0$, **B** influx $j_B$, **A** escape flux $f \cdot J_A$, and an empirical EZ length $\lambda$. Note that there is no empirical mechanism to estimate the fraction $f$ itself; hence the value of the actual incident species **A** areal flux is unknown. Nonetheless, the crucial quantity in the model is the fractional escape flux itself. The simple, yet effective model provides valuable insight into the underlying growth mechanism, and an empirical mechanism to estimate growth parameters. The extracted values of the fluxes can be plotted as a function of the growth durations or the annulus outer diameter for the same growth, as shown in Figure 4d. The basic thickness $d_0$ and the estimated EZ lengths from the fits can also be plotted, as shown in Figure 4e.

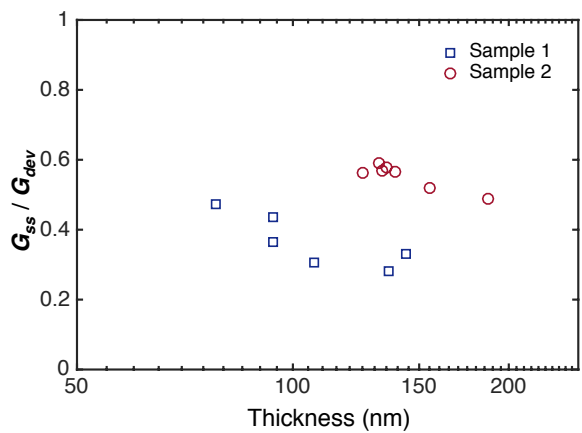

**Figure S11.** Ratio of the TSS conductance to the total device sheet conductance for CF-vdWE grown TI Hall bars as a function of the Hall bar thickness from fits to an effective two-channel conduction model (Equation 2 in main text),[S5,S13] indicating TSS-dominated transport at room-temperature.

## Supporting Information References